\documentclass[prx,reprint,superscriptaddress, nofootinbib, floatfix, longbibliography]{revtex4-2}
\usepackage[utf8]{inputenc}

\usepackage[dvipsnames]{xcolor}

\usepackage{amsmath}
\usepackage{amssymb}
\usepackage{textcomp}
\usepackage{longtable}
\usepackage{graphics}
\usepackage{gensymb} 
\usepackage{graphicx}
\usepackage{epsfig}
\usepackage{epstopdf}
\usepackage{soul}
\usepackage{xcolor}
\usepackage{float}
\usepackage{hyperref}

\usepackage{physics}
\usepackage{changepage}

\begin{document}

\title{Coexisting Ballistic and Diffusive Heat Transport \\in Micrometer-Long Molecular Junctions}

\author{P. M. Martinez}
\affiliation{Instituto de Ciencia de Materiales de Madrid (ICMM), CSIC, Madrid, Spain}
\affiliation{Departamento de F\'{i}sica Te\'{o}rica de la Materia
                  Condensada, Universidad Aut\'{o}noma de Madrid,
                  E-28049 Madrid, Spain}
\author{O. Mateos-Lopez}
\affiliation{Instituto de Ciencia de Materiales de Madrid (ICMM), CSIC, Madrid, Spain}
\affiliation{Departamento de F\'{i}sica Te\'{o}rica de la Materia
                  Condensada, Universidad Aut\'{o}noma de Madrid,
                  E-28049 Madrid, Spain}
\author{J. C. Cuevas}
\email{juancarlos.cuevas@uam.es}
\affiliation{Departamento de F\'{i}sica Te\'{o}rica de la Materia
                  Condensada, Universidad Aut\'{o}noma de Madrid,
                  E-28049 Madrid, Spain}
\affiliation{Condensed Matter Physics Center (IFIMAC),
                     Universidad Aut\'{o}noma de Madrid, E-28049 Madrid, Spain}
\author{J. G. Vilhena}
\email{guilherme.vilhena@csic.es}
\affiliation{Instituto de Ciencia de Materiales de Madrid (ICMM), CSIC, Madrid, Spain}




%
\begin{abstract}
\noindent Boltzmann transport theory, the standard framework for predicting thermal conductivity, assumes that every vibrational mode eventually scatters, acquiring a finite lifetime that yields a convergent, length-independent thermal conductivity: Fourier's law.
Here we show that this assumption fails in a real molecular system.
Through atomistic simulations of Au--alkane--Au single-molecule junctions spanning five orders of magnitude in length (0.5~nm to 4~$\mu$m), we find that thermal conductivity never converges.
Transport is ballistic for up to one hundred nanometers at room temperature, extending nearly two orders of magnitude beyond existing single-molecule measurements.
Past this window, conductivity diverges as $L^{1/3}$, the scaling predicted by the Kardar-Parisi-Zhang universality class for momentum-conserving systems.
Frequency-resolved decomposition of the heat current reveals the mechanism behind the divergence.
Low-frequency acoustic modes never thermalize: protected by momentum conservation, they remain ballistic at every chain length, still carrying 50\% of the total heat current at $L = 2~\mu$m.
All other modes thermalize collectively as discrete vibrational states merge into scattering-active phonon bands with increasing length.
Hence, the diverging conductivity emerges from the boundary between these coexisting transport regimes: as $L$ grows, the onset of scattering shifts progressively toward lower frequencies, suppressing the ballistic channel at a rate that sustains the $L^{1/3}$ divergence, leaving a finite contribution at every length.
This coexistence of permanent ballistic and well-behaved diffusive transport, anticipated in abstract one-dimensional lattice models, survives the structural and chemical complexity of real micrometer-sized junctions.
\end{abstract}

\date{This manuscript was compiled on \today}

\maketitle

\noindent Heat transport in low-dimensional materials holds great technological potential owing to their unique properties~\cite{Segal2003,Henry2008,Cui2019,Chen2021}.
As early as 1955, the {\it gedanken experiment} of Fermi, Pasta, Ulam, and Tsingou (FPUT)~\cite{Fermi1955} on non-linear 1D lattices revealed a striking anomaly: energy injected into such systems remained circulating across a few vibrational modes, showing no tendency to spread evenly or to reach equipartition.
Yet, the standard frameworks for predicting heat transport rely on the opposite assumption.
Boltzmann transport equation (BTE), the standard microscopic framework for predicting heat conduction, rests on the basic assumption that lattice vibrations scatter on short timescales compared to their transit across the sample, ensuring local thermal equilibrium.
Under this premise, the local heat flux ($\phi$, energy flow per unit time and area) is strictly proportional to the imposed temperature gradient ($\grad T$),
\begin{equation}\label{eq:Fourier}
    \phi = -\kappa_{\rm th} \grad T,
\end{equation}
\noindent defining the thermal conductivity, $\kappa_{\rm th}$, as a finite, length-independent material property.
The FPUT directly challenges this picture: if energy never thermalizes across the modes, the foundational premises of both the BTE and Fourier's law no longer hold in one dimension.
The possibility of enhancing $\kappa_{\rm th}$ beyond its reference bulk values \cite{Yang2021, Shen2010} has profound repercussions on thermal management applications, potentially bringing enhanced performance for heat dissipation and transport.
Exploiting the unique properties of non-Fourier transport would facilitate unprecedented control of heat flows through the development of novel devices such as thermal rectifiers \cite{Lepri2003} or heat-to-work thermoelectrics \cite{Benenti2023}.

Recent research suggests a general law for the divergence of $\kappa_{\rm th}$ in non-Fourier regimes.
Derivations employing non-linear fluctuating hydrodynamics predict $\kappa^{\rm 1D}_{\rm th}\propto L^{1/3}$ and link it to the Kardar-Parisi-Zhang (KPZ) universality class \cite{Spohn2014,Barreto2019}, a claim supported by numerous simulations \cite{Mai2007,Chen2013,Dematteis2020} and recent experiments \cite{Yang2021}.
The universality of the exponent, beyond any chemical properties of the sample, suggests a fundamental connection between dimensionality and energy diffusion \cite{Nayaran2002,Li2003,Chen2013}, although its ubiquity in real materials remains an open debate in the research community.
The forecast of a general, specific exponent is grounded on integrals of motion or conserved magnitudes such that, in the asymptotic limit, specific microscopic details of the interaction model fade away.
Similar deviations have also been predicted in 2D materials \cite{Nayaran2002}, although following a logarithmic $\kappa^{\rm 2D}_{\rm th}\propto \ln(L)$ divergence owing to their distinct dimensionality.
This range of non-Fourier scalings is broadly known as superdiffusion, as opposed to traditional Fourier diffusion where $\kappa_{\rm th}$ is constant.
Crucially, these predictions are largely derived from abstract one-dimensional lattice models~\cite{Nayaran2002,Chen2013,Dematteis2020,Benenti2023}; whether they survive the structural and chemical complexity of real materials remains an open question.

Notwithstanding, direct validation of such theoretical forecasts remains a formidable challenge.
Most experiments are subject to boundary scattering effects, which make direct observation of non-Fourier transport difficult.
In this regard, the environment imposes a maximum length for vibrations to propagate inside the material that, if sufficiently small, promotes ballistic heat transport \cite{Segal2016,Klockner2016}.
In such regime, vibrations traverse the medium unperturbed, linking hot and cold spots without suffering energy losses in their journey.
Those critical lengths rapidly decrease with growing temperature, restricting most ballistic transport reports to cryogenic conditions.
However, recent single-molecule experiments performed on short, nanometer-sized organic polymers \cite{Cui2019, Mosso2019, Yelishala2025} show that ballistic transport can be sustained even at room temperature.
In such ballistic regime, $\kappa_{\rm th}$ grows linearly with $L$ as a trivial consequence of length-independent conductance, rather than from an anomaly in the scattering physics.
Direct observation of the crossover from ballistic to superdiffusive transport in a chemically realistic system, as opposed to an abstract lattice model, has to our knowledge not yet been reported.

\begin{figure*}[t]
    \centering
    \includegraphics[width=\textwidth]{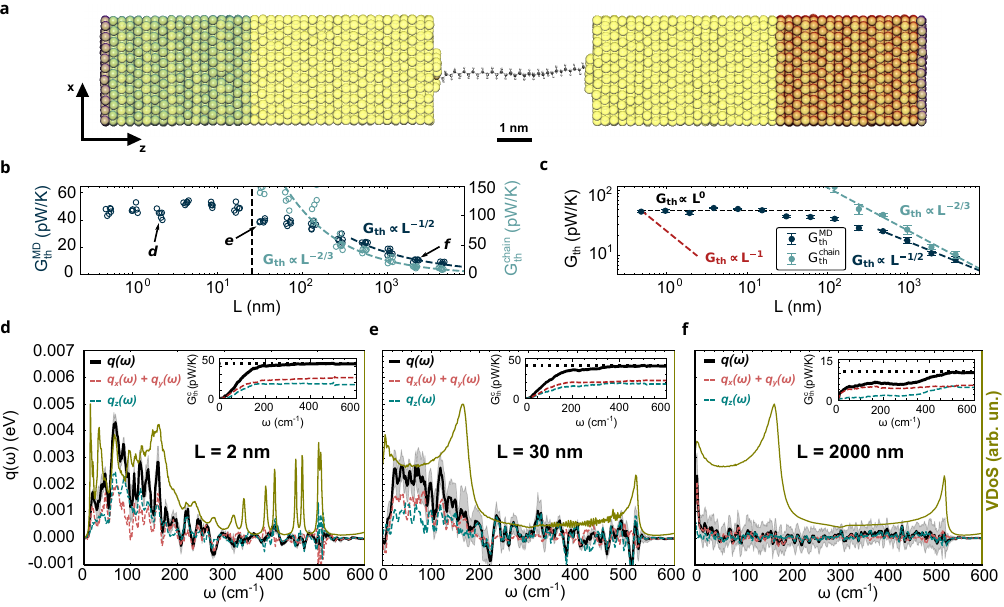}
    \caption{\textbf{Ballistic-to-superdiffusive crossover and its spectral fingerprint in Au-alkane-Au junctions.}
    \textbf{a)} Representative Au-alkane-Au junction (310~K, 0.35~nN tension). Shading denotes harmonic restraints (gray) and thermal baths (green: 290~K, orange: 330~K).
    \textbf{b)} Total thermal conductance $G^{\rm MD}_{\rm th}$ (navy, left axis) and intrinsic chain conductance $G^{\rm chain}_{\rm th}$ (cyan, right axis) as a function of chain length. Point dispersion is introduced for visualization purposes.
    \textbf{c)} Log-log plot of b). Dashed lines guide different scaling regimes: ballistic ($L^0$), superdiffusive ($L^{-2/3}$ or $L^{-1/2}$), and diffusive ($L^{-1}$). Error bars: standard deviation of 6 runs.
    \textbf{d)} Vibrational density of states (green) and spectral heat current $q(\omega)$ (black) for a 2~nm chain, split into longitudinal (teal) and transverse (pink) components. Gray shading: standard deviation of $q(\omega)$ across 10 spatial planes. Inset: Cumulative spectral conductance converging to the $G^{\rm MD}_{\rm th}$ total value.
    \textbf{e, f)} Same as (d) for 30-nm and 2000-nm chains, showing the progressive suppression of high-frequency ballistic channels.}
    \label{fig:main}
\end{figure*}

In this work, we bridge this gap by demonstrating the room-temperature crossover from ballistic to superdiffusive heat transport in chemically-accurate single-molecule junctions.
Through atomistic non-equilibrium molecular dynamics (NEMD) simulations, we predict ballistic heat transport in Au-alkane-Au molecular junctions up to 20 nm under moderate tension and up to 100 nm at higher pulling forces, followed by a length scaling consistent with $\kappa^{\rm 1D}_{\rm th} \propto L^{1/3}$.
Our results agree with state-of-the-art experiments in the short-chain limit \cite{Cui2019, Mosso2019} and are consistent with measurements in other 1D materials~\cite{Yang2021}, reinforcing the universality of the exponent.
Convergence to the diffusive scaling is absent even for 4-micrometer chains, the longest wires under consideration.
The spectral decomposition of the heat current \cite{Saaskilahti2014, Fan2017} shows that only low-frequency acoustic modes are efficient energy carriers at any length.
The breakdown of ballistic transport is linked to the participation of higher-frequency longitudinal modes whose contribution is initially negligible.
This transition toward superdiffusion can be delayed by increasing the tension, thus preserving ballistic transport up to $L=100\,$nm in highly stretched chains.
Crucially, long-wavelength vibrations remain ballistic regardless of chain length, even after the macroscopic crossover to superdiffusion.
Their contribution amounts to roughly half of the total thermal conductance at $L \sim 2~\mu$m, positioning them as the microscopic drivers of the anomalous scaling.
The remaining frequencies thermalize simultaneously upon increasing $L$, as discrete vibrational states transition to phonon bands that facilitate internal scattering.
These observations are at odds with the central assumption of Boltzmann transport theory, where each mode acquires a finite scattering lifetime, and show that Fourier's law fails to describe heat transport in real single-molecule junctions even at room temperature.

\section*{Ballistic-to-superdiffusive crossover and long-chain limit}

\noindent We show the simulation setup in Fig.~\ref{fig:main}a, where an alkane chain of variable length contacts two gold electrodes.
These are connected to external reservoirs that maintain a temperature difference $\Delta T = 40~\text{K}$, thereby driving a heat current $\dot{Q}$ across the bridging molecule.
Meanwhile, the energy exchanged with the thermostats per unit time is recorded, which equals $\dot{Q}$ by virtue of energy conservation (see Methods for details).
Finally, the thermal conductance is obtained as $G_{\rm th}^{\rm MD} = \dot{Q} / \Delta T$. The connection to the usual coefficient $\kappa_{\rm th}^{\rm MD}$ is straightforward; by definition $G_{\rm th}^{\rm MD}= \kappa_{\rm th}^{\rm MD}L^{-1}A$, where $A$ is the cross-section of the sample.

Unlike prior approaches~\cite{Henry2008,Wang2017}, our methodology does not rely on presumptions regarding the transport regime nor suffer from finite samplings; all molecular vibrations are fully accounted for at every chain length $L$.
Moreover, finite-temperature NEMD benefits from sampling all energetically accessible conformations and is not restricted to linear response theory.
Therefore, it constitutes an accurate and unbiased framework to model one-dimensional heat transport~\cite{Yang2021,Cui2019,Mosso2019,Yelishala2025}, all while retaining spectroscopic accuracy -- see SI Appendix.

We performed at least 6 independent NEMD simulations to estimate $G_{\rm th}^{\rm MD}$ for 14 chain lengths ranging from $0.5 \,$nm to $4 \, \mu$m. 
We find that $G_{\rm th}^{\rm MD}$ remains constant up to 20~nm-long chains as shown in Fig.~\ref{fig:main}b.
The predicted value $G_{\rm th}^{\rm MD}\sim 48 \,$pW/K for the plateau agrees with previous non-equilibrium Green's function calculations \cite{Klockner2016} ($G_{\rm th}\sim 38 \,$pW/K) and other NEMD \cite{Wang2022} ($G_{\rm th}\sim 45 \,$pW/K) simulations, all above single-molecule experiments \cite{Mosso2019, Cui2019}.
An exact quantitative agreement relies on atomically-precise couplings \cite{Losego2012,Yelishala2025} which are irrelevant to the transport regime.
In essence, the interfacial regions greatly influence the effective contact resistance and determine the value of the ballistic plateau.

Beyond this critical length, $G_{\rm th}^{\rm MD}$ monotonically decreases with $L$. To gauge whether this decay complies with Fourier's Law $\left(G_{\rm th}\sim L^{-1}\right)$, we present the results on logarithmic scale in Fig.~\ref{fig:main}c. 
The last 4 points follow a superdiffusive scaling 
$G_{\rm th}^{\rm MD}\sim L^{-0.5}$ 
deviating from Fourier's Law. These chains map $0.5$ $\mu$m$<L<4.1$ $\mu$m, showing that the diffusive picture is invalid even on the micrometer scale.

Nevertheless, the long-chain trend is still influenced by the interface contact resistance.
We estimate this contribution through a simple 3-resistor model to isolate the inherent scaling of the chain -- see SI Appendix for details. 
The intrinsic chain conductance $G_{\rm th}^{\rm chain}$ is shown in Fig.~\ref{fig:main}b-c in cyan, alongside the original $G_{\rm th}^{\rm MD}$ in navy.
In the long-chain limit, both conductances converge once the interfacial contact resistance becomes negligible compared to the total, that is, the interface plus the chain itself.
By subtracting this interfacial contribution, we get $G_{\rm th}^{\rm chain}\sim L^{-0.66\pm0.08}$ in agreement with the universal law proposed for 1D systems and recently found in van-der-Waals crystal nanowires \cite{Yang2021}.
We note that new experimental works on alkane-grafted nanoparticle melts report an exponent of $-1/2$ \cite{Liu2024}. Such complex samples, however, incorporate a large number of interfacial resistances that complicate their analysis without relying on coarse approximations. On the contrary, the contact resistance of the suspended nanowires \cite{Yang2021} is estimated to be negligible, providing a cleaner framework to study the evolution with length.
Based on our results, ballistic transport persists for at least 20~nm in suspended Au-alkane-Au junctions, with no recovery of Fourier scaling up to $L = 4~\mu$m.

\section*{Spectral fingerprint of heat transport in alkane chains}

\noindent To gain insight into the source of the transition, we decompose the heat current into its frequency-resolved components~\cite{Saaskilahti2014, Fan2015, Fan2017}. We define the spectral heat current $q(\omega)$ as:

\begin{equation}
 \dot{Q} = \int_0^\infty q(\omega) d\omega,
\end{equation}

\noindent
which assesses the power carried by a given frequency $\omega$ in the steady state. 
All terms of the interaction potential, including anharmonic ones, are explicitly included
in the definition of $q(\omega)$, which ensures that the connection between $\dot{Q}$ and $q(\omega)$ is exact -- see SI Appendix. 
Furthermore, it has been shown \cite{Fan2015} that this definition is equivalent to the formal microscopic heat flux operator \cite{Irving1950,Hardy1963,Hardy1982} in NEMD simulations.

We present the results of such decomposition in Fig.~\ref{fig:main}d-f for various chains to show their contributions to overall transport.
The vibrational density of states (VDoS, green) indicates which molecular modes are populated.
Whether a vibrational mode is occupied does not necessarily mean that it channels a net energy flux $q(\omega)$ between both electrodes.
Such contribution is represented in black, whose integral yields the total heat flow through the junction $\dot{Q}$.
The heat current spectrum $q(\omega)$ can be further decomposed into transverse (pink) or longitudinal (blue) motions with respect to the main axis of the molecule.
Each of the lower panels depicts how transport occurs at 3 distinct lengths, indicated with labels d-f in Fig.~\ref{fig:main}b.

In short chains (Fig.~\ref{fig:main}d) only those modes with $\omega<200 \,$cm$^{-1}$ contribute meaningfully to the overall energy transfer. 
Beyond such cutoff frequency, the net energy current $q(\omega)$ fluctuates around zero due to numerical noise.
This is observed in the cumulative integral of $G_{\rm th}$, $G^{\rm c}_{\rm th}(\omega) = \Delta T^{-1}\int_0^\omega q(x)dx$, which saturates at the value $G_{\rm th}^{\rm MD}$ -- see inset of Fig.~\ref{fig:main}d. 
This low frequency cutoff coincides with the Debye frequency of the gold electrodes ($\sim\,160\,$cm$^{-1}$), which illustrates their function as low-pass filters for transport. 
The short chain length ensures that no internal scattering occurs, thus preserving the initial filtering introduced by the lead chemistry.

The absence of scattering can be verified by computing $q(\omega)$ at different spatial regions along the chain, which yields identical signals up to numerical noise -- see shaded regions of Fig.~\ref{fig:main}d-f for standard deviations across 10 planes. 
This position-invariance of $q(\omega)$ constitutes a stronger test of ballisticity than previously available: while single-molecule experiments \cite{Cui2019, Mosso2019} infer ballistic transport from the length-independence of $G_{\rm th}$, a necessary but not sufficient condition, and other theoretical calculations \cite{Klockner2016} assume elastic transmission by construction, the spatially-resolved spectral current directly proves that each frequency channel traverses the junction without internal scattering.
Changes in electrode chemistry would enable probing vibrations populated beyond 200 cm$^{-1}$, which bounce back and forth at the interfaces with zero contribution to energy transport \cite{Klockner2016}.

In sufficiently long chains, elastic transport breaks down. 
The spectral heat current $q(\omega)$ becomes position-dependent, as evidenced by its standard deviation (shaded in Fig. \ref{fig:main}e).
Specifically, the $\omega<200\,$cm$^{-1}$ range exhibits the largest variance, signaling that internal scattering primarily affects these low-frequency modes.
Nevertheless, the integral extended to all $\omega$ -- see inset -- remains similar to those of ballistic chains even at $L=30 \,$nm, meaning that chain-driven resistive processes are weak and do not perturb the net heat flux noticeably.

The overall thermal resistance increases with the length as these resistive events accumulate.
This is due to a larger likelihood of transferring some energy to back-propagating waves.
We observe the decay in Fig. \ref{fig:main}f, where all contributions to the thermal conductance are suppressed except for long-wavelength modes.
Note that the phonon density of states (in green) indicates that plenty of modes are populated, but they do not drive a net energy flux (in black).
Interestingly, a new contribution emerges at $\omega\sim400 \,$cm$^{-1}$ accounting for up to half of the total $G_{\rm th}$ -- see Fig. \ref{fig:main}f inset -- which lies beyond the Debye frequency of gold.
Therefore, it can only be attributed to internal scattering inside the molecular chain.
In essence, the impact of the low-pass filter weakens as the length increases and scattering due to the inherent chemistry of the bridging wire arises.
Further decomposition of $q(\omega)$ into transverse and longitudinal polarizations shows that long-wavelength vibrations are essentially transverse to the main molecular axis.
We relate these low-frequency features to the non-attenuating transverse and torsional modes previously identified by Henry and Chen in periodic polymer chains \cite{Henry2008}. Because higher-frequency modes scatter and hence undergo conventional Fourier diffusion, it is precisely the persistence of these non-decaying transverse channels that dictates the anomalous length dependence of the macroscopic heat flux. By spectrally isolating their contribution to the total heat flux in a chemically realistic molecular junction, our results provide the first direct link between the non-attenuating modes reported in model polymer chains and the macroscopic divergence of $\kappa_{\rm th}$ observed in non-Fourier transport.

\section*{Delay of the superdiffusive onset through mechanical pulling}
\noindent The distinct spectral roles of transverse and longitudinal polarizations motivate studying heat transport under varying tension, which restricts the available conformational space and preferentially stiffens the transverse channel.
We find that increasing the tension delays the regime crossover, effectively extending the conductance plateau to $L=100 \,$nm. 
This preserves lossless heat flow over distances nearly two orders of magnitude beyond current experimental records for single-molecule junctions \cite{Cui2019,Mosso2019}.

\begin{figure}[htbp]
    \centering
    \includegraphics[width=0.48\textwidth]{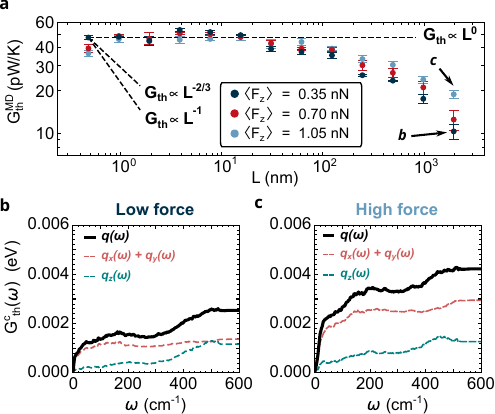}
    \caption{\textbf{Enhancement of ballistic transport through mechanical pulling.} 
    \textbf{a)} Thermal conductance versus chain length under various applied tensions (averaged across 3 independent NEMD runs). Dashed lines indicate theoretical scalings for ballistic, superdiffusive, and diffusive regimes. Letters denote the specific systems detailed in (b) and (c). 
    \textbf{b)} Cumulative thermal conductance for a representative $2\,\mu$m chain under low tension ($\langle F_z \rangle = 0.35$~nN). The total conductance is decomposed in transverse (pink) and longitudinal (teal) components. 
    \textbf{c)} Same as (b) for a highly stretched chain ($\langle F_z \rangle = 1.05$~nN), showing the delayed onset of superdiffusion.}
    \label{fig:forces}
\end{figure}

We compare the effect of the tensile strain for 3 force values as represented in Fig.~\ref{fig:forces}a.
Remarkably, $G^{\rm MD}_{\rm th}$ doubles for $\sim$2-$\mu$m-long chains compared to the low-tension results.
Close inspection of $q(\omega)$ for the chains labeled in Fig.~\ref{fig:forces}a reveals the spectral origin of this enhancement.
The longitudinal contribution (teal) to the heat flux is nearly identical for both pulling forces (Fig.~\ref{fig:forces}b-c).
The major difference lies in the transverse channel (pink): at the lowest frequencies, highly-stretched chains conduct heat manifestly better, with the transverse contribution roughly doubling between the low- and high-tension cases. This is consistent with the identification of transverse modes as the persistent ballistic channel: tension stiffens these modes, enhancing their capacity to carry heat over long distances.
Previous works had linked low-strain conformations to lower values of $G_{\rm th}$ when gauche defects appear \cite{Li2015}, but the lowest pulling force under study already yields stretched conformations. Therefore, defect formation is not responsible for the conductance difference shown in Fig.~\ref{fig:forces}.
Moreover, localized defects should be essentially transparent for low-frequency modes. Rather, the overall stiffening of the chain promotes efficient heat transport, as observed experimentally for 1D nanowires \cite{Yang2021}.

Changes in tension are akin to changes in pressure, which can select distinct universality classes for the anomalous exponent~\cite{Spohn2014,Benenti2023, Lepri2016_b}. In particular, the nonzero tension in our system places it in the regime where the KPZ exponent $1/3$ is theoretically predicted~\cite{Spohn2014,Benenti2023, Lepri2016_b}.
Although a confident assessment of the exponent's dependence on tension would require additional simulations for longer chains, the results shown in Fig.~\ref{fig:forces} allow us to conclude that Fourier's law is never recovered at micrometer-sized chains regardless of the applied force.

\section*{Thermalization and ballistic low-energy modes}

\begin{figure}[htbp]
    \centering
    \includegraphics[width=0.385\textwidth]{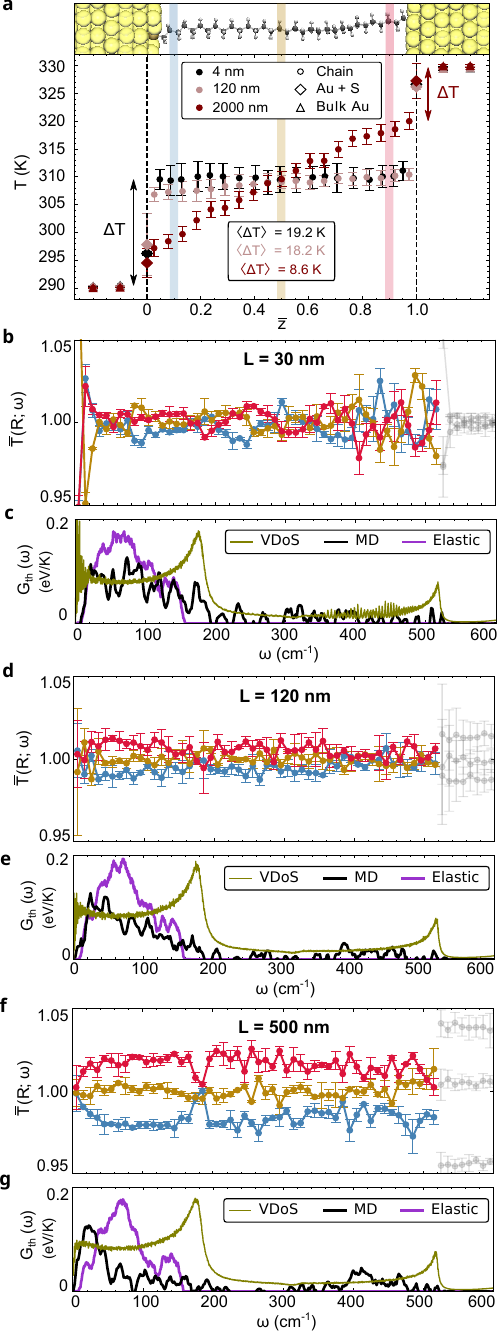}
    \caption{\textbf{Chain thermalization and persistent non-interacting modes in Au-alkane-Au junctions.} 
    \textbf{a)} Temperature profiles versus normalized coordinate ($z/L$) for varying chain lengths. $\Delta T$ is the boundary temperature jump.
    \textbf{b)} Normalized spectral temperature $\bar{T}(\omega)$ for a 30~nm chain at $0.1L$ (blue), $0.5L$ (yellow), and $0.9L$ (red). Grey points denote modes contributing $<0.5\%$ to the total kinetic energy. Error bars: standard deviation over three 20~ns runs. 
    \textbf{c)} Vibrational density of states (green) and spectral heat current $q(\omega)$ (black) for the 30~nm chain, compared with the purely elastic limit (purple). 
    \textbf{d--g)} As in (b,c) for 120~nm (d,e) and 500~nm (f,g) chains. The spatial splitting of $\bar{T}(\omega)$ reveals the onset of thermalization.}
    \label{fig:coexistence}
\end{figure}

\noindent The presence of an inelastic contribution to $q(\omega)$ beyond the Debye frequency of gold proves that intrachain scattering is present.
Let us consider that heat transfers ballistically, that is, vibrations enter and leave the chain unperturbed.
This means that the kinetic energy and, thus, temperature are not functions of space.
Fig. \ref{fig:coexistence}a shows the average temperature profiles for 3 independent simulations of highly stretched chains as a function of the normalized position.
Shorter polymer chains, represented in black, showcase a flat temperature profile as they offer zero opposition to the energy flow. 
All the relevant resistance lies at the organometallic interface, hence the large temperature jumps reflected in Fig. \ref{fig:coexistence}a.

There is a tight relationship between the breakdown of ballistic transport and the reduction of this interfacial temperature jump. 
As the chain length increases, both ends of the chain appear at different temperatures, as shown in Fig. \ref{fig:coexistence}a.
This means that the edges of the wires are vibrationally distinct; internal scattering has appeared and impacts heat conduction.
To understand which modes are thermalizing the chain, we extend the frequency analysis and define a local, frequency-dependent temperature:
\begin{equation}
T(\mathcal{R}; \omega_1,\omega_2) 
= \frac{1}{3 k_B N_\mathcal{R}}
\sum_{i\in \mathcal{R}} m_i 
\int_{\omega_1}^{\omega_2}
\abs{ \tilde{\mathbf{v}}_i(\omega) }^2 \, d\omega,
\end{equation}
\noindent where $\tilde{\mathbf{v}}_i(\omega)$ is the Fourier transform of the velocity of atom $i$, $m_i$ its mass, and $\abs{\tilde{\mathbf{v}}_i(\omega)}^2$ is normalized such that 
 $\langle \abs{\mathbf{v}_i(t)}^2 \rangle_t = \int_0^\infty \abs{\tilde{\mathbf{v}}_i(\omega)}^2 d\omega$. 
The sum is performed for all $N_\mathcal{R}$ atoms within a selected region $\mathcal{R}$. 
This constitutes a measure of the kinetic energy stored in frequencies $\omega\in[\omega_1,\omega_2]$ inside a spatial region $\mathcal{R}$. Since not all frequencies are homogeneously populated, we normalize this magnitude as:
\begin{equation}
  \bar{T}\left(\mathcal{R}; \frac{\omega_1+\omega_2}{2}\right) = \frac{nT(\mathcal{R}; \omega_1,\omega_2)}{\sum_i^nT(\mathcal{R}_i; \omega_1,\omega_2)} ,
\end{equation}
\noindent with $n$ being the number of regions probed. This normalizes $T(\mathcal{R}; \omega_1,\omega_2)$ with respect to its mean value along the chain. This is repeated for each interval $[\omega_1,\omega_2]$ as a function of the average frequency $\frac{\omega_1+\omega_2}{2}$.
Thus, $\bar{T}(\mathcal{R}; \omega)<1$ means region $\mathcal{R}$ has a lower temperature than the chain average and conversely if $\bar{T}(\mathcal{R}; \omega)>1$.
We calculate $\bar{T}(\mathcal{R}; \omega)$ at 3 spatial regions ($0.1L$, $0.5L$ and $0.9L$) for a 30~nm-long chain in Fig. \ref{fig:coexistence}b.

All spectral temperatures oscillate around unity, showing that the vibrational state is statistically equivalent anywhere in the chain for any frequency.
We also compare the NEMD conductance to the strict elastic limit (see Methods), in which energy exchanges across modes are forbidden (Fig.~\ref{fig:coexistence}c).
Despite minor differences, the two spectra are broadly consistent.
This is further proof that heat transport is indeed ballistic for short chains.

When the same analysis is performed in longer chains, specifically at the regime crossover (Fig. \ref{fig:coexistence}d), we find the inception of mode thermalization.
The spectral temperature transitions from the previous $\bar{T}(\mathcal{R}; \omega)=1$ independent of $\omega$ and $\mathcal{R}$ to a position-dependent distribution.
In particular, this process starts in the same frequency range where the NEMD spectrum deviates from the elastic limit -- see Fig. \ref{fig:coexistence}e.

Strikingly, most frequencies appear to thermalize simultaneously as a function of L.
The simultaneous thermalization is more apparent for the 500~nm chain shown in Fig.~\ref{fig:coexistence}f-g.
Critically, only the long-wavelength limit remains ballistic while the remaining populated vibrations reach a constant $\bar{T}$.
This collective thermalization differs from the progressive, frequency-by-frequency relaxation expected in a phonon gas picture. While Dematteis et al.~\cite{Dematteis2020} showed that FPUT lattices develop a coexistence of ballistic and kinetic sectors, in our system the kinetic sector thermalizes collectively rather than through a cascade from high to low frequencies.
In our system, which includes the full anharmonic coupling between all modes, most frequencies thermalize at the same chain length -- see SI Appendix.

\section*{Breakdown of kinetic theory and divergent thermal conductivity in quasi-one-dimensional systems}
\noindent The spectral decomposition of the heat current reveals two coexisting transport regimes in superdiffusive chains ($L > 100$~nm): persistent ballistic channels, dominated by low-frequency modes, and a well-behaved diffusive sector comprising frequencies above a length-dependent critical threshold ($\omega > \omega_c(L)$, as shown by the thermalized frequencies in Fig.~\ref{fig:coexistence}f). 
These ballistic channels carry over 50\% of the total thermal conductance at $L \sim 2~\mu$m and show no sign of thermalization across the five orders of magnitude in length here considered while the diffusive sector thermalizes collectively.
It is the interplay and progressive shift of this $\omega_c(L)$ boundary, rather than the isolated behavior of any single mode, that produces the divergent $\kappa_{\rm th}$.

Independent confirmation of this spectral bifurcation is provided by comparing the full NEMD conductance to the purely elastic Landauer-B\"{u}ttiker (LB) limit~\cite{Klockner2016}.
The LB formalism enforces coherent, scattering-free propagation by construction, thus isolating the purely ballistic contribution to transport.
As shown in Fig.~S2, the elastic conductance $G_{\rm th}^{\rm LB}$ matches the NEMD plateau up to $L \approx 100$~nm, but increasingly overestimates the unbiased macroscopic $G_{\rm th}^{\rm MD}$ beyond this threshold.
This growing divergence between the two conductances at longer chain lengths directly quantifies the relevance of inelastic scattering within the diffusive sector.
Hence, neither the LB nor BTE alone can describe the full physics: the former captures the ballistic channels correctly but cannot account for the thermalized modes, while the latter~\cite{Wang2017} assumes finite scattering lifetimes for modes that our simulations show remain strictly ballistic.
Only the full anharmonic dynamics of NEMD naturally captures the coexistence of both regimes that governs heat transport in these systems.

We argue that the persistence of these ballistic channels is not a chemical peculiarity of alkane junctions, but a fundamental kinematic consequence of quasi-one-dimensional geometry.
In any momentum-conserving 1D system, long-wavelength acoustic modes are inherently protected because such energy and momentum conservation constrain the available scattering phase space as $\omega \to 0$~\cite{Lepri2003,Lepri2016_b,Spohn2014}.
For transverse branches specifically, a quadratic long-wavelength dispersion~\cite{Wang2017,Lindsay2010} strongly suppresses intra-branch three-phonon processes. 
While standard kinetic models predict these modes are heavily damped by inter-branch cross-couplings~\cite{Wang2017}, our observations reveal that these transverse channels remain protected and fail to thermalize across five orders of magnitude in length.
Physically, this occurs because such lowest-frequency vibrations are fully delocalized across the entire length of the structure ($\lambda \sim L$). 
Governed by the macroscopic bending stiffness of the chain, they should be blind to the local atomic details that would otherwise mediate such anharmonic scattering.
The chemical insensitivity is corroborated explicitly by computing the fundamental normal modes for three structurally diverse linear polymers: aliphatic, conjugated, and fused-ring backbones (Fig.~S8).
Despite substantial differences in local orbital hybridization and molecular rigidity, the lowest-frequency vibrations invariably manifest as global transverse bending motions.
This geometric protection explains why independent studies have observed similar non-decaying modes in isolated polymer chains using entirely different theoretical models and force fields~\cite{Lepri2000,Chen2013,Crnjar2018}, such as the pioneering work of Henry and Chen~\cite{Henry2008}, providing strong evidence that these non-thermalizing channels are a generic structural feature of 1D molecular architectures.

This ballistic-diffusive coexistence also provides a microscopic physical origin of the KPZ universality class in heat transport.
The anomalous exponent $\kappa \propto L^{1/3}$ is not merely a fitting parameter; it reflects the rate at which the critical boundary $\omega_c(L)$ between the ballistic and diffusive sectors migrates.
The total thermal conductance comprises a Fourier-compliant diffusive term ($G_{\rm diff} \propto 1/L$) and a ballistic contribution ($G_{\rm bal}$) governed by the Landauer-B\"{u}ttiker limit for frequencies below $\omega_c(L)$.
As the chain lengthens, $\omega_c$ shifts to lower frequencies (see Fig.~\ref{fig:main}), causing $G_{\rm bal}$ to decrease as formerly protected modes enter the diffusive sector. 
Crucially, our data shows that this erosion proceeds at such a slow rate that the ballistic channel is never fully exhausted. 
Instead of the conventional $1/L$ decay, this persistent ballistic contribution forces the total conductance to decay anomalously slowly as $G_{\rm th} \propto L^{-2/3}$, resulting in a divergent $\kappa_{\rm th} \propto L^{1/3}$.
Because the rate of this $\omega_c$ shift is governed by momentum conservation and phase-space symmetries rather than specific interatomic potentials~\cite{Spohn2014,Nayaran2002,Benenti2023}, the scaling exponent is expected to be universal across 1D momentum-conserving systems. 
This generality is consistent with the experimental observation of $\kappa_{\rm th} \propto L^{1/3}$ in NbSe$_3$ nanowires~\cite{Yang2021} -- a system fundamentally distinct from alkanes in both composition and interatomic interactions.

These findings pose a direct challenge to the phonon gas interpretation of heat flow.
In the standard kinetic picture, heat is carried by discrete wavepackets with well-defined mean free paths $\Lambda(\omega)$ and relaxation times $\tau(\omega)$, where every mode eventually thermalizes through independent scattering events.
This formalism, embodied in the linearized BTE (LBTE), assigns finite lifetimes to all modes and thus guarantees a finite $\kappa_{\rm th}$ by construction~\cite{Mcgaughey2006,Simoncelli2022}.
Our results contradict this premise.
For the modes that our spectral decomposition identifies as persistently ballistic, the assumption of finite relaxation times finds no support in the simulation data.
Recovering Fourier’s law in these systems would require that the low frequency-modes eventually thermalize; yet we observe no evidence of such thermalization across five orders of magnitude in length.
Furthermore, the underlying physics of momentum-conserving scattering in 1D provides no clear mechanism for such thermalization at finite lengths.
For instance, first-principles LBTE calculations on single polyethylene chains~\cite{Wang2017} predict the convergence of $\kappa_{\rm th}$ at $\sim$1 mm as normal processes activate resistive scattering of the lowest-frequency modes, thus indirectly suppressing the lifetimes of transverse modes.
Our results suggest that this indirect thermalization is absent under the full anharmonic dynamics captured by NEMD, pointing to a fundamental limitation of the perturbative treatment of anharmonicity employed in standard kinetic frameworks.

\section*{Conclusions and outlook}

\noindent We have demonstrated the full room-temperature crossover from ballistic to superdiffusive heat transport in Au-alkane-Au single-molecule junctions spanning $0.5\,$nm to $4\, \mu$m.
Ballistic conduction persists for at least $20\,$nm under ambient tension and up to $100\,$nm under higher pulling forces, extending the ballistic regime nearly two orders of magnitude beyond current single-molecule measurements~\cite{Cui2019,Mosso2019,Yelishala2025}.
In the long-chain limit, the intrinsic chain conductance follows $G_{\rm th}^{\rm chain} \sim L^{-0.66\pm0.08}$, consistent with the $\kappa \propto L^{1/3}$ scaling of the KPZ universality class and with recent experiments on van der Waals crystal nanowires~\cite{Yang2021}.
No recovery of Fourier scaling is observed up to $L = 4~\mu$m.

The spectral decomposition of the heat current identifies the microscopic origin of this divergence: low-frequency transverse modes that never thermalize, carrying over half of the total heat current even at micrometer scales.
These non-thermalizing phonons are the microscopic drivers of the anomalous scaling.
Their persistence should not be specific to alkane chemistry: equivalent global bending modes appear in conjugated and fused-ring polymers (Fig.~S8). Non-decaying autocorrelations for transverse and torsional modes have been independently reported in infinite polyethylene chains~\cite{Henry2008}.
While the critical length for the ballistic-to-superdiffusive crossover will vary with the specific material and temperature, the existence of a persistent ballistic channel appears to be an intrinsic feature of quasi-one-dimensional momentum-conserving systems~\cite{Chen2021,Benenti2023}.

At all experimentally accessible lengths, from the nanometer-scale junctions measured by Cui et al.~\cite{Cui2019} and Mosso et al.~\cite{Mosso2019} to the micrometer-scale wires studied here, thermal transport is governed by the interplay of ballistic and diffusive sectors rather than by a single, length-independent $\kappa_{\rm th}$.
This has direct consequences for thermal modeling in molecular electronics, polymer nanofibers, and any device architecture where heat flows through effectively one-dimensional channels.
Future work combining the spectral framework developed here with emerging experimental techniques capable of probing thermal transport in aligned polymer fibers~\cite{Shen2010}, nanowires~\cite{Yang2021}, and single-molecule junctions~\cite{Cui2019,Yelishala2025} should enable a direct experimental verification of the ballistic-diffusive coexistence and its role in setting the anomalous thermal properties of quasi-one-dimensional materials.

\section*{Materials and Methods}
{\small
\noindent All MD simulations were performed using LAMMPS v23Jun2022 \cite{LAMMPS} or GROMACS v2023 \cite{GROMACS}. 
We used a velocity Verlet integrator in LAMMPS and leap-frog stochastic-dynamics integrator in GROMACS, both with a timestep of $0.25\,$fs.  GROMACS was used only on the longest $L\sim 4\,\mu$m chains to accelerate the pulling. All equilibrium simulations were performed in either Gibbs (NPT) or Canonical (NVT) ensemble, with a constant number of atoms ($N_{\rm Au} =$11,200 and a length-dependent number for the chains) and temperature ($T=310\,$K).We used periodic boundary conditions (PBC) and a $1\,$nm real-space cutoff for Van der Waals interactions, ensuring no interchain PBC interactions. Temperatures and thermostats energies were recorded every $25$~ps. Barostats and thermostats implement Berendsen and Langevin algorithms with damping rates of 2 and 100 ps respectively.

We use the embedded atom method for gold \cite{Sheng2011}, the Morse potential between Au and terminal S \cite{Mahaffy1997} and OPLS-AA parameters \cite{Jorgensen2005, LigParGen} for the chain, with Lorenz-Berthelot rules for the gold-chain cross-interactions \cite{HeinzLJ}. We neglect the Coulomb terms owing to the highly stretched chain conformations which we verified does not compromise the accuracy of molecular vibrations -- see SI Appendix.

The simulation protocol comprises pressure equilibration, NVT constant force pulling, and heat-flux NEMD. In the first stage, we enforce p=1 bar along the transverse (xy) directions as temperature increases from 1 K with random velocity assignment to 310 K. During the  constant-force pulling, Au boundary layers are harmonically restrained ($k_{xy}$= 1 N m$^{-1}$ and $k_z$= 10 N m$^{-1}$ per atom) excepting the Z direction of the topmost layer to allow the pulling, which is done for at least 10 ns. This continues until the end to end distance plateaus, taking the last 5-ns time-average distance for the production run (fixed with $k_z=10\,$N m$^{-1}$).
For $L\sim 4\,\mu$m, the pulling was carried in GROMACS for the isolated chain. The final conformation is placed between the electrodes reservoirs and equilibrated in the NVT ensemble in LAMMPS before production. Finally, electrode ends are driven to $290\,$K and $330\,$K respectively for $0.5\,$ns. NEMD productions run for $40\,$ns while maintaining the temperature bias and discarding the first $20\,$ns. Heat current $\dot{Q}$ is quantified as the slope of a linear fit of total energy pumped or withdrawn with respect to time. Simulations are extended as long as necessary until the difference between slopes is smaller than $|\dot{Q}_{in}-\dot{Q}_{out}|\Delta T^{-1}< 1\,$pW/K.

The VDOS is computed from the modulus squared of the Fourier Transform of the molecule’s atomic velocities~\cite{Lee1993}. The spectral decomposition of the heat current $q(\omega)$ \cite{Saaskilahti2014} is computed in a subsequent 30 ns long MD run. We express frequencies in wavenumbers as usual units of molecular vibrations ($\omega = 33.36$ cm$^{-1}=1\, \mbox{THz} = 4.136 \, \mbox{meV}$). Transmissions $\tau$ and elastic spectral conductances are computed with Green's functions techniques \cite{Klockner2018} up to 165 cm$^{-1}=20.5\, \mbox{meV}$, beyond which transmissions become negligible. Dynamical matrices are extracted from LAMMPS through a FIRE minimization~\cite{Guenole2020} of selected contact geometries with the same potential as for the NEMD simulations. Absorbing boundary conditions with a phenomenological broadening of $0.5\, \mbox{THz}$ are applied to simulate the effect of an infinite reservoir \cite{Romero2021}. The Landauer-B\"{u}ttiker spectral conductances are obtained as $G_{th}^{LB}(\omega) = {C_V(\omega) \tau(\omega)}$, where $C_V(\omega)$ is the heat capacity of a Bose gas.}

\section*{Acknowledgments}
\small{
We thank O. Gutierrez-Varela for fruitful discussions and G. Prampolini, J. Cerezo and A. Semmeq for their assistance assessing OPLS-AA FF accuracy.
We acknowledge Stefan Bilan and computing resources at Picasso, Finisterrae3 (RES-FI-2023-0031, RES-FI-2023-2-0006).
P.M.M. acknowledges the Spanish Ministry of Education and Professional Formation (Grant No. FPU21/06224).
J.C.C. thanks the Spanish Ministry of Science and Innovation (PID2024-157536NB-C22) and the "Maria de Maeztu" Units of Excellence in R\&D (CEX2023-001316-M).
J.G.V. acknowledges funding from the Spanish CM “Talento Program” (Project No. 2020-T1/ND-20306), the Spanish Ministerio de Ciencia e Innovación (Grant Nos. PID2020-113722RJ-I00, TED2021-132219A-I00, CNS2023-144011 and PID2024-157536NB-C21), and the Severo Ochoa Centres of Excellence (CEX2024-001445-S). }

\small{P.M.M, J.C.C and J.G.V. designed research; P.M.M. performed research; P.M.M. and O.M.L. analyzed data; and P.M.M., J.C.C. and J.G.V. wrote the paper.}


\bibliographystyle{apsrev4-2}
\bibliography{main,SI}

\clearpage
\onecolumngrid

\section*{SI Appendix}

\setcounter{figure}{0} 
\setcounter{table}{0}  
\setcounter{section}{0} 
\setcounter{equation}{0} 

\renewcommand{\thefigure}{S\arabic{figure}}
\renewcommand{\thetable}{S\arabic{table}}
\renewcommand{\thesection}{S\arabic{section}}
\renewcommand{\theequation}{S\arabic{equation}}

\subsection*{NEMD spectral heat current}

\noindent Due to the many-body nature of the OPLS potential, we follow the most general expression \cite{Fan2017} for the microscopic heat flux operator to compute the energy current across a control surface separating sides $A$ and $B$, following the derivation in \cite{Saaskilahti2014}:

\begin{equation}
 q(\omega) = -2\Re{\int C(t) e^{-i\omega t} dt},
\end{equation}

\noindent
with $C(t)$ the time correlation function:
\begin{equation}\label{eq:Kij}
C(t) = -\sum_{i,j \in A,B} \left[\frac{\partial U_{i}}{\partial \mathbf{r_{ij}}}(0)\mathbf{v_j}(t) - \frac{\partial U_{j}}{\partial \mathbf{r_{ji}}}(0)\mathbf{v_i}(t)\right].
\end{equation}

Sampled values of positions and velocities from the NEMD dynamics every 3.75 fs are used as inputs of \ref{eq:Kij} for the calculation of $q(\omega)$. This sampling is sufficient to capture the largest populated frequency $\sim 3100$ cm$^{-1}$ (C-H stretching). The analytical derivatives are computed from the intrachain interaction force-field (OPLS-AA):

\begin{equation}\label{eq:OPLS}
 U(\{r_{ij}\}) \sum_s^{N_s} k^{s}(r-r_0)^2 + \sum_b^{N_b} k^{b}(\theta-\theta_0)^2 + \sum_d^{N_d} \sum_{n=1}^4 k_{n}^d (1 +\cos{n\phi}) + \sum_i^{N_p}\sum_{j\neq i}^{N_p} 4\epsilon_{ij}\left(\frac{\sigma_{ij}^{12}}{r_{ij}^{12}}-\frac{\sigma_{ij}^6}{r_{ij}^6}\right),
\end{equation}

\noindent
where we note that it can be conveniently expressed as a function of the set of displacement vectors $\{r_{ij}\}$. The energy exchange is computed across a C-C backbone bond for 10 planes spanning the entire molecular length, observing similar values for the total integral of all planes and consistent ($<1\%$ relative error) with the value tallied from the thermostats, thus ensuring current conservation. Due to their limited influence on distant atoms, non-bonded terms further than 3 CH$_2$ units from the control surface were neglected in the calculation of $q(\omega)$. This approximation is ultimately justified by the good agreement between the integral of $G^c_{\rm th}(\omega)$ and the stationary value $G^{\rm MD}_{\rm th}$ obtained from NEMD -- see insets in Fig.~\ref{fig:main}d-f of the main text. All spectra output are convoluted with a rectangular function of width 0.25 THz to smooth out statistical noise as in the original publications \cite{Saaskilahti2014, Saaskilahti2015}.

\subsection*{Landauer-B\"{u}ttiker framework}

\noindent The transmission $\tau$ in the Landauer-B\"{u}ttiker framework is defined as in \cite{Klockner2018}:

\begin{equation}
    \tau(E) = Tr\{ D^r_{CC}\Lambda_L{D^r_{CC}}^\dagger\Lambda_R\},
\end{equation}

\noindent where $D^r_{CC}$ is the retarded Green’s function of the central part:

\begin{equation}
    D^r_{CC} = \left[(E+i\eta)^2\mathbb{I}-K_{CC}-\Sigma_R^r-\Sigma_L^r\right]^{-1}
\end{equation}

\noindent and $\Lambda_Z,Z\in\{L,R\}$ are the line-broadening matrices:

\begin{equation}
    \Lambda_Z= i[\Sigma_Z^r-{\Sigma_Z^r}^\dagger].
\end{equation}

We have defined our central region as all atoms composing the chains and terraces as well as the nearest 3 gold layers. The left and right regions are defined as the following 4 gold layers in all cases (620 atoms per electrode) The embedding self-energies $\Sigma_Z^r=K_{CZd^r_{ZZ}K_{ZC}}$ are defined in terms of each reservoir retarded Green’s functions:

\begin{equation}
    d^r_{ZZ} = \left[(E+i\eta_F)^2\mathbb{I}-K_{ZZ}\right]^{-1},
\end{equation}

\noindent which should mimic an infinite system coupled to the central region through the $K_{CZ}$ terms of the dynamical matrix:

\begin{equation}
K=
\begin{pmatrix}
    K_{LL} & K_{LC} & 0\\
    K_{CL} & K_{CC} & K_{CR}\\
    0 & K_{LC} & K_{RR}\\
\end{pmatrix}.
\end{equation}

Note that the $L,R$ reservoirs are not directly coupled, a condition that is met owing to the short-range character of the interatomic potentials. To reproduce the effect of an infinite system with the finite-rank matrices , we set absorbing boundary conditions with a finite broadening $\eta_F=0.5$ THz rather than the infinitesimally small $\eta=10^{-6}$ THz. This introduces a phenomenological broadening in the density of states, such that a finite collection of Dirac deltas can mimic the effect of an infinite system coupled to the central region. We show in Supplementary Fig.~\ref{fig:SI:absorbing} the resulting local density of states (LDOS):

\begin{equation}
    LDOS _{ZZ}= -2\ Tr\{E\pi^{-1}\Im\{{d_{ZZ}}\}\}
\end{equation}

\noindent compared to the finite-temperature (T = 300 K) bulk VDoS recorded for 1 ns in bulk gold. Both curves have been normalized.

\begin{figure}[H]
    \centering
    \includegraphics[width=0.5\textwidth]{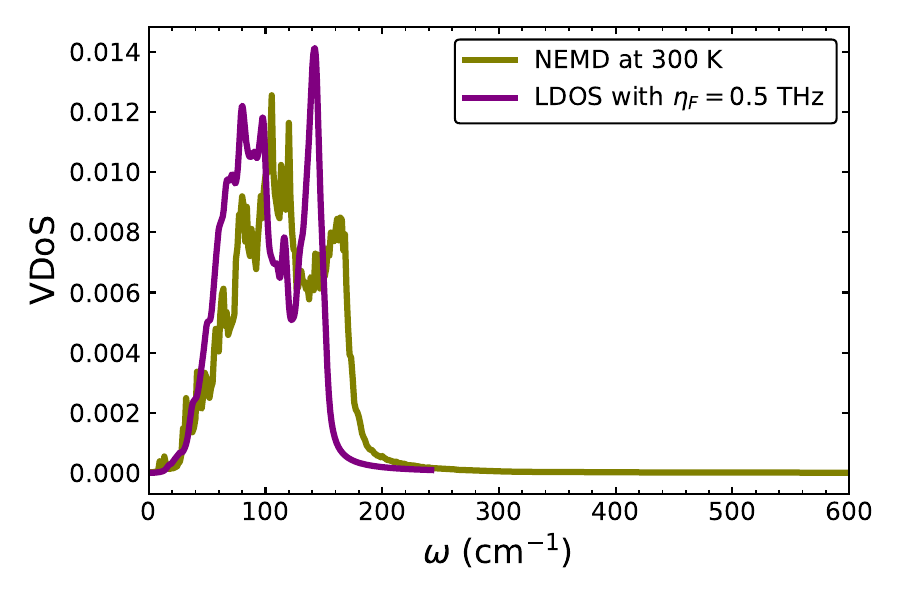}
    \caption{\textbf{Numerical treatment of semi-infinite reservoirs.} The imposed broadening $\eta_F=0.5$ THz mimics the effect of a semi-infinite gold reservoir in contact with the central part.}
    \label{fig:SI:absorbing}
\end{figure}

\subsection*{Transition from coherent to incoherent transport}

\noindent The breakdown of the plateau found for $G_{\rm th}^{\rm MD}$ may in principle be due to elastic or inelastic scatterings. We have analyzed the length dependence of the conductance within the LB formalism, which enforces elastic energy propagation. Both methods yield similar phonon thermal conductances up to 100 nm. Elastic transport cannot account for the phonon conductance decay beyond this point, as we find $G_{\rm th}^{\rm LB}(500\,$nm$)\sim 40\,$pW/K versus $G_{\rm th}^{\rm LB}(500 \,$nm$)\sim 25\,$pW/K.

Therefore, we validate that the decay is due to phonon-phonon interactions within the molecular chain.

\begin{figure}[H]
    \centering
    \includegraphics[width=0.65\textwidth]{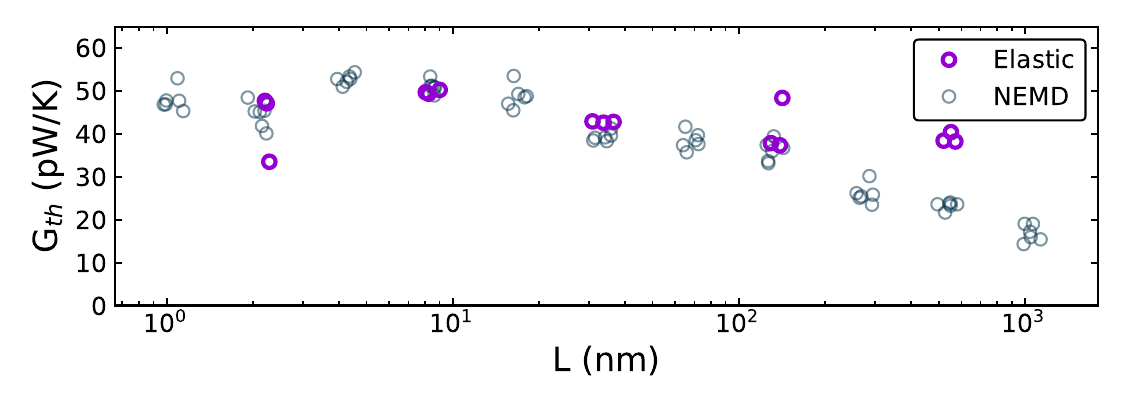}
    \caption{\textbf{Elastic conductance as a function of length.} NEMD thermal conductances of Fig. \ref{fig:main}b (navy) and Landauer-B\"{u}ttiker conductances in the elastic limit (purple).}
    \label{fig:SI:STD_SHC}
\end{figure}

\subsection*{Frequency-resolved kinetic energy contributions}

\noindent Spectral temperatures $T(\mathcal{R};\omega)$ are presented as ratios with respect to their position-averaged values $\bar T(\mathcal{R};\omega)$. For this reason, not all points appearing in Fig.~\ref{fig:coexistence} of the main text have the same weight in the thermal profile. In particular, for those regions corresponding to a phonon bandgap, as $550$ cm$^{-1}<\omega<650$ cm$^{-1}$, we find ratios slightly higher but with no contribution to transport and they are of numerical origin.

\begin{figure}[H]
    \centering
    \includegraphics[width=0.5\textwidth]{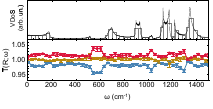}
    \caption{\textbf{Thermalization of optical modes.} Normalized temperature for a 20-ns NEMD run as in Fig.~\ref{fig:coexistence}f. The upper panel shows high-resolution VDoS and a lower resolution binning matching the sampling as in the lower panel.}
    \label{fig:SI:SKE}
\end{figure}

These vibrations are not defined within a local harmonic approximation, meaning that no phonon or vibrational states can be properly associated to them. Fig.~\ref{fig:SI:SKE} extends the thermalization analysis to the nearest optical bands. The ratio of kinetic energies is also approximately constant at high frequencies showcasing full thermalization.

\subsection*{Reported chain lengths across different pulling forces}

\noindent In the main text, we discuss our results on the basis of chain lengths. In practice, we generate geometries that successively duplicate the number of monomers of the previous sample, ranging from $2^2$ CH$_2$ monomers up to $2^{15}$ monomers. We estimate an average length of 0.12 nm per monomer for the extended chain, thus obtaining the reported lengths, which match reasonably well the NEMD end-to-end distances. The exact lengths are subject to small variations due to statistical fluctuations and pulling strengths. However, the relative change between the final conformations after the 0.35 nN and 1.05 nN pullings for a $2^{14} \sim 2$~$\mu$m chain is less than 2.5\%, which in our view justifies a direct comparison of thermal conductances across various pulling regimes.

\subsection*{Thermostats and finite size effects}

\begin{figure}[H]
    \centering
    \includegraphics[width=0.6\textwidth]{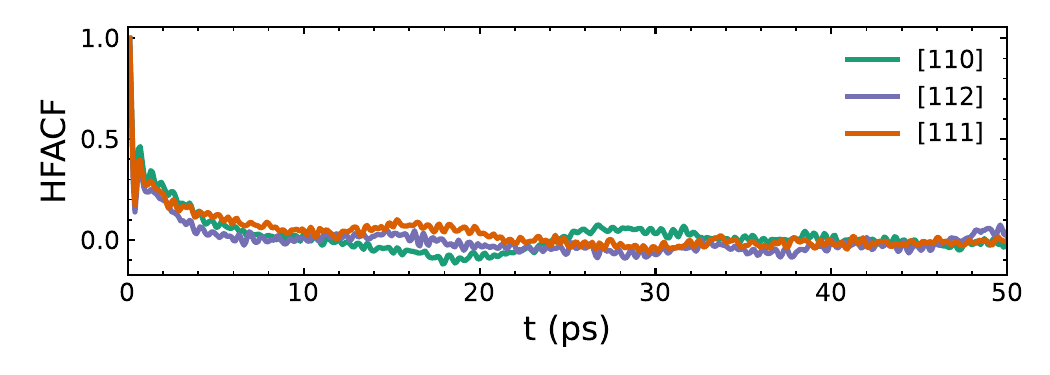}
    \caption{\textbf{Autocorrelation of the heat flux in gold in an NVE run.} Bulk sample consists of a 5x5x5 nm = 2,100 atoms cube, whose cross section matches the electrodes used in the NEMD simulations. Assuming an exponential decay, time constants are 1.2 ps, 0.74 and 0.97 ps for the $[1\bar{1}0]$, $[11\bar{2}]$ and $[111]$ crystallographic directions respectively. The thermostat coupling used ($\tau= 2.0$ ps) does not have a big impact on the natural decay of vibrations.}
    \label{fig:SI:autocorrelation}
\end{figure}

\noindent Thermostat time constants and reservoir sizes have been determined following the recommendations of \cite{OlartePlata2022} and \cite{Li2019} respectively. The heat flux autocorrelation function in an NVE simulation for bulk gold, with $N=2,100$, $<T>=310$ K and $<p>=1$ bar is shown in Fig.~\ref{fig:SI:autocorrelation}. The decay is fitted to an exponential assuming a single relaxation time. The time constant was found to be $\sim1.2$ ps.

Considering the speed of sound in gold $\sim3,200$ m/s, this natural relaxation time requires at least 4 nm of thermostated gold to fully ensure phonon relaxation. In our study, we use 4 nm of thermostated gold in addition to 6 nm of unthermostated gold near the interface. Despite the relatively large distance, there is no thermal gradient formed inside the electrodes. This is proof that the details concerning the thermostat scheme should not influence our results and that heat through the chain is driven by two bodies with temperature difference $\Delta T=40$ K.

\subsection*{Binding site and sulfur motion}

\begin{figure}[H]
    \centering
    \includegraphics[width=0.32\textwidth]{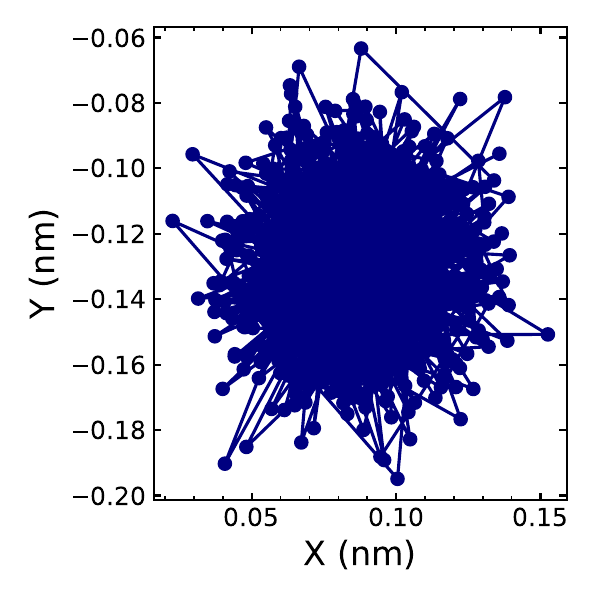}
    \caption{\textbf{Sulfur diffusion during production.} Trajectory followed by one sulfur atom during a sample 30~ns production run in LAMMPS for $L = 120$ nm and $<F_z>=1.05\,$nN. The sulfur position oscillates around its adsorption site.}
    \label{fig:SI:S_motion}
\end{figure}

\noindent The adequacy of the in-vacuo GROMACS pulling rests on the fixed position of the terminal S atoms found during any LAMMPS simulation; see Fig.~\ref{fig:SI:S_motion}. Thus, we may neglect the electrode and adsorbate dynamics and only simulate the end-to-end pulling in GROMACS for the longest chain. After equilibration, the chain is relocated to its preferred adsorption site and relaxed in an NVT LAMMPS simulation before the generation of the thermal bias.

We note that sulfur atoms diffuse on a pristine surface. The fact that molecules are pinned relies on the terrace-like geometry that we selected for the study. This should best mimic experimental conditions as the molecule is firmly attached to both electrodes.

\subsection*{Force field accuracy}

\noindent We employ the OPLS-AA interaction model~\cite{Jorgensen2005} ignoring point charges, which are crucial in condensed phases but which we find have negligible impact under vacuum conditions. For simple molecular species like alkanes, this simplification retains quantum-mechanical accuracy  when compared to its gas-phase DFT Hessian.

\begin{figure}[H]
    \centering
    \includegraphics[width=.5\textwidth]{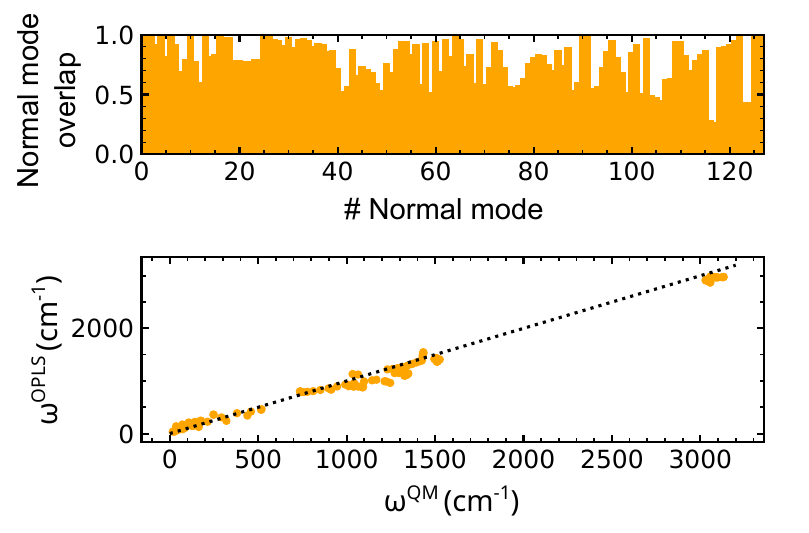}
    \caption{\textbf{Validation of the OPLS-AA force field against independent QM-DFT data.} Comparison of the Hessian matrix calculated at DFT level with the OPLS-AA classical potential in tetradecane. Top panel corresponds to the inner product between eigenvectors, where the pair asignment is done based on the maximum overlap starting from high frequencies towards low frequencies. Bottom panel shows the correlation between the corresponding eigenfrequencies.}
    \label{fig:SI:QM}
\end{figure}

The reference Hessian was obtained with the B3LYP exchange correlation functional~\cite{LYP1988,Becke1993}, cc-pvdz basis set and Becke-Johnson D3 dispersion correction~\cite{BJ2011} with Gaussian09~\cite{Gaussian09}. In the upper panel, we represent the inner product of the classical and DFT eigenmodes, reaching 1 when the mapping is perfect and 0 if they were orthogonal. Note that specifically, low-frequency modes map the first principles reference. The lower panel shows the correlation between the Hessian eigenfrequencies in the classical and quantum-mechanical frameworks.

\subsection*{Chain conductance from a 3-resistor model}

\noindent Consider the thermal circuit composed of 3 resistances in series. The first and last resistances represent the thermal resistance of the gold reservoirs and the organometallic junction opposing to the energy flow from hot to cold regions. We will name their sum, that is, the total contact resistance of the system, $\left(G_{\rm th}^{\rm bal}\right)^{-1}$. We may write the total resistance probed in the NEMD simulations as:
 
\begin{equation}
 R_{\rm th} = 2R_{\rm th}^{contact} + R_{\rm th}^{chain} = \left(G_{\rm th}^{\rm bal}\right)^{-1} + \eta_\alpha L ^{\alpha},
\end{equation}

\noindent where we consider as fitting parameters $\left(G_{th}^{bal}\right)^{-1},\eta_\alpha$ and $\alpha$ as the contact resistance, a proportionality constant, and the length scaling of the conductance, respectively. We perform a non-linear fit using the Levenberg-Marquad algorithm, taking the inverse of the squared NEMD uncertainties as weights for the least-square fit.

\begin{figure}[H]
    \centering
    \includegraphics[width=0.65\linewidth]{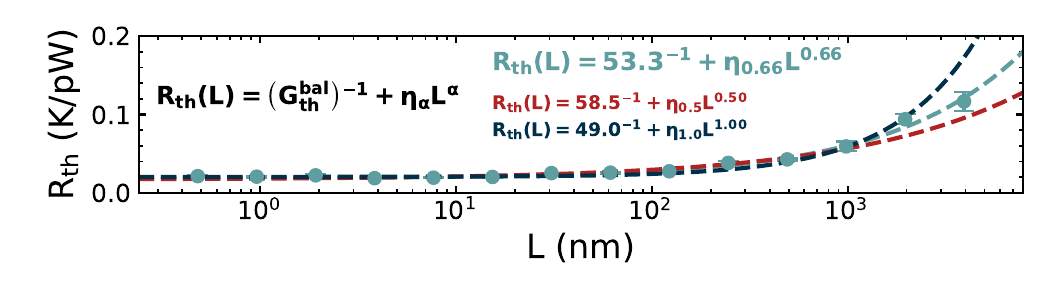}
    \caption{\textbf{Determination of superdiffusive exponent and contact resistance.} Thermal resistance from NEMD simulations and best fit (cyan). Best fits enforcing scaling exponents 1/2 and 1 are also shown for reference.}
    \label{fig:SI:3resistor}
\end{figure}

The best fit for an exponent is $\alpha=0.66\pm0.08$, while the zero-length extrapolated conductance is $G_{\rm th}^{\rm bal}=53\pm2\,$pW/K, in good agreement with the plateau found in NEMD.

\subsection*{Universality of superdiffusion in 1D systems}

\noindent In the main text, we argue that our results are not limited to the molecular species under study. Alkanes represent the simplest and most homogeneous backbone of stable organic compounds. We find that the anomalous behavior is linked to the lack of thermalization of long-wavelength modes. The character of these normal modes is not affected by the specific chemistry of the backbone. The Hessians of other linear polymers are obtained with Gaussian09~\cite{Gaussian09}, using the B3LYP exchange-correlation functional, cc-pvdz basis set and Becke-Johnson D3 dispersion correction. The fundamental normal modes involve global motions of the chain regardless of the conjugation scheme. 

\begin{figure}[h]
    \centering
    \includegraphics[width=0.5\textwidth]{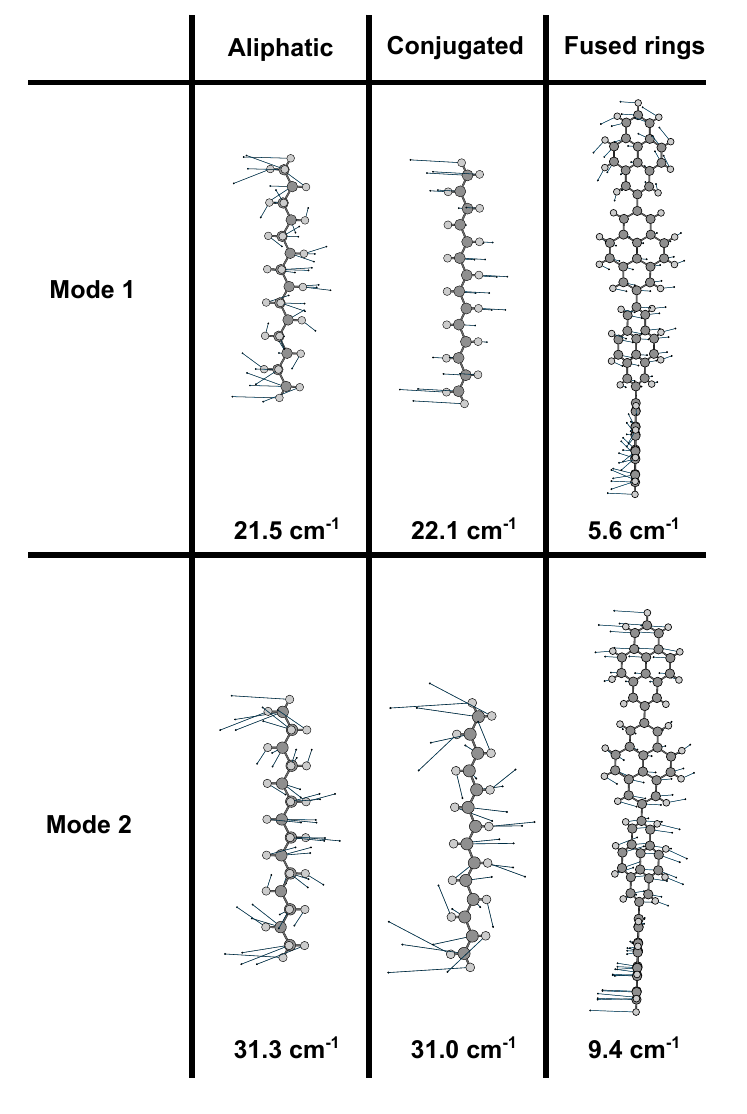}
    \caption{\textbf{First two normal modes for different hybridization schemes.} The two lowest frequency modes are shown for aliphatic (tetradecane), conjugated (tetradeca-1,3,5,7,9,11,13-heptaene) and fused-ring monomers (2,7-pyrenylene tetramer).}
    \label{fig:SI:hybridmols}
\end{figure}

Consequently, we expect that superdiffusion is ubiquitous in any 1D system. Even if the breakdown of ballistic conduction in each case may differ owing to different phonon-phonon scattering for higher frequency modes, the low-frequency protected vibrations are shared by any 1D chain. 

\end{document}